\documentclass[12pt]{article}
\usepackage{amsmath,amssymb,bm,graphicx}
\usepackage{color} 

\begin{document}

\begin{flushright}
\end{flushright}

\vskip 0.5 truecm

\begin{center}
{\Large{\bf A path integral derivation of the equations of anomalous Hall effect
}}
\end{center}
\vskip .5 truecm
\centerline{\bf Kazuo Fujikawa~$^1$
 {\rm and} Koichiro Umetsu~$^2$ }
\vskip .4 truecm
\centerline {\it $^1$~Interdisciplinary Theoretical and Mathematical Sciences Program (iTHEMS), 
}
\centerline {\it   RIKEN, Wako 351-0198, 
Japan}
\vskip 0.4 truecm
\centerline {\it $^2$~Laboratory of Physics, College of Science and Technology, and Junior College, }
\centerline{\it Funabashi Campus, Nihon University, Funabashi, Chiba 274-8501, Japan}
\vskip 0.5 truecm

\makeatletter
\@addtoreset{equation}{section}
\def\theequation{\thesection.\arabic{equation}}
\makeatother

\begin{abstract}

 A path integral (Lagrangian formalism) is used to  derive the effective equations of motion of the anomalous Hall effect with Berry's phase on the basis of the adiabatic condition $|E_{n\pm1}-E_{n}|\gg 2\pi\hbar/T$, where $T$ is the typical time scale of the slower system and $E_{n}$ is the energy level of the fast system.  
  In the conventional definition of the adiabatic condition with $T\rightarrow {\rm large}$ and fixed energy eigenvalues, no commutation relations are defined for slower variables by the Bjorken-Johnson-Low prescription except for the starting canonical commutators.   On the other hand, in a singular limit $|E_{n\pm1}-E_{n}|\rightarrow \infty$ with specific $E_{n}$ kept fixed for which any motions of the slower variables $X_{k}$ can be treated to be adiabatic, the non-canonical dynamical system with deformed commutators and the Nernst effect appear. 
In the Born-Oppenheimer approximation based on the canonical commutation relations, the equations of motion of the anomalous Hall effect is obtained if one uses an auxiliary variable $X_{k}^{(n)}=X_{k}+{\cal A}^{(n)}_{k}$ with Berry's connection ${\cal A}^{(n)}_{k}$ in the absence of the electromagnetic vector potential $eA_{k}(X)$ and thus without the Nernst effect. It is shown that the gauge symmetries associated with Berry's connection and the electromagnetic vector potential $eA_{k}(X)$ are incompatible in the canonical Hamiltonian formalism.  
The appearance of the non-canonical dynamical system with the Nernst effect is a consequence of the deformation of the quantum principle to incorporate the two incompatible gauge symmetries.

\end{abstract}

\section{Introduction}

Historically, the geometric phase or Berry's phase \cite{Berry} has been recognized as a forgotten phase of the state vector in the naive applications of the adiabatic theorem. The topological aspects of Berry's phase such as a Dirac monopole-like structure in the level crossing problem  in the precise adiabatic limit are well-known \cite{Simon}. The use of  Berry's phase in the analyses of the anomalous Hall effect \cite{Karplus, Fang} and the anomalous spin Hall effect \cite{Hirsch}  
 has, however, revealed  the more general aspects of Berry's phase, namely, a dynamical system which is not quantized in the conventional canonical formalism of quantum mechanics \cite{Niu} leading to deformed commutators. The related non-commutative geometry  has been known for many years if one chooses  suitable variables in the Born-Oppenheimer approximation which is by itself quantized by the canonical commutation relations \cite{Blount, Sinitsyn}. 

If one understands Berry's phase as a characteristic property of the quantum mechanical system in the adiabatic limit \cite{Born, Kato}, it is puzzling why Berry's phase leads to a non-canonical system which deforms the canonical commutation relations dictated by the principle of quantum mechanics \footnote{In the present article, Berry's phase is considered only in the context of quantum mechanics.}. The purpose of the present article is to clarify the puzzling aspect by presenting a systematic derivation of the equations of motion of the anomalous Hall effect by the  path integral (Lagrangian) formalism, starting with the quantum system defined by the standard canonical commutation relations. We use the second quantized formulation of Berry's phase which makes it transparent that Berry's phase appears when one throws away some terms causing the level crossing in  the adiabatic limit \cite{deguchi}, rather than the common view that Berry's phase is an additional term in the adiabatic limit.

To be explicit, 
the effective equations of motion of the anomalous Hall effect, which 
incorporate Berry's phase  near the level crossing point, are customarily adopted as (see, for example,  \cite{Niu})
\begin{eqnarray}\label{semi-classical equation}
\dot{x}_{k}=-\Omega_{kl}(\vec{p})\dot{p}_{l} +\frac{\partial \epsilon_{n}(\vec{p})}{\partial p_{k}}, \ \
\dot{p}_{k}=-eF_{kl}(\vec{x})\dot{x}_{l} + e\frac{\partial}{\partial x_{k}}\phi(\vec{x})
\end{eqnarray}
by adding the adiabatic Berry's phase $\Omega_{kl}\dot{p}_{l}$ to the equations of motion as an extra induced term. Here $\epsilon_{n}(\vec{p})$ essentially  stands for the n-th energy level of the fast system.  The magnetic flux $\Omega_{kl}(\vec{p})$ of Berry's phase, which is  assumed to be a {\em genuine Dirac monopole form}~\cite{Dirac} for the moment, and the electromagnetic tensor $F_{kl}$ are defined by  
\begin{eqnarray}
\Omega_{kl}=\frac{\partial}{\partial p_{k}}{\cal A}_{l}-\frac{\partial}{\partial p_{l}}{\cal A}_{k}, \ \  \ F_{kl}=\frac{\partial}{\partial x_{k}}A_{l}-\frac{\partial}{\partial x_{l}} A_{k},
\end{eqnarray}
respectively. The monopole is located at the origin of momentum space, i.e., at the level crossing point in the adiabatic level-crossing problem, but it is now assumed to be a genuine particle in the momentum space.  Here we defined $p_{l}=\hbar k_{l}$ to write all the equations in terms of $p_{l}$ 
compared to the notation in \cite{Niu}, to keep track of the $\hbar$ factor in a transparent way.

It is known that  the equations in \eqref{semi-classical equation} are derived without using any commutation relations from~\cite{Duval}  
\begin{eqnarray}\label{action-1}
S=\int dt[p_{k}\dot{x}_{k} - eA_{k}(\vec{x}, t)\dot{x}_{k} + {\cal A}_{k}(\vec{p})\dot{p}_{k}-\epsilon_{n}(\vec{p})+e\phi(\vec{x})],
\end{eqnarray}  
A simplified version of this action was used in \cite{Niu}.  It is usually assumed that the action \eqref{action-1} with Berry's phase included is a fundamental {\em classical action}, and it has been shown~\cite{Duval} that the action \eqref{action-1}, which is known to belong to a {\em non-canonical system} \cite{Niu}, is treated by a modified canonical formalism inverting a symplectic matrix defined by the action in the extended phase space formalism \cite{Faddeev-Jackiw}. They then derived the deformed Poisson brackets induced by the genuine monopole curvature $\Omega_{kl}=\epsilon^{klm}\Omega_{m}$,
\begin{eqnarray}\label{Poisson bracket} 
&&\{x_{k},x_{l}\}=\frac{\epsilon^{klm}\Omega_{m}}{1+e\vec{B}\cdot\vec{\Omega}} , \ \ \ \{p_{k},x_{l}\}=-\frac{\delta_{kl}+e\Omega_{k}B_{l}}{1+e\vec{B}\cdot\vec{\Omega}},\nonumber\\
&&\{p_{k},p_{l}\}=- \frac{\epsilon^{klm}eB_{m}}{1+e\vec{B}\cdot\vec{\Omega}},
\end{eqnarray}
where the factors containing $\Omega_{k}$ are anomalous \cite{Duval}.  The background field method applied to \eqref{action-1} in the path integral formalism using the Bjorken-Johnson-Low (BJL) prescription~\cite{BJL}, which is briefly summarized in Appendix,  gives the commutation relations \cite{fujikawa-prd2018}
\begin{eqnarray}\label{commutator-1} 
&&[x_{k},x_{l}]= i\hbar\frac{\epsilon^{klm}\Omega_{m}}{1+e\vec{B}\cdot\vec{\Omega}} , \ \ \ [p_{k},x_{l}]=- i\hbar\frac{\delta_{kl}+e\Omega_{k}B_{l}}{1+e\vec{B}\cdot\vec{\Omega}},\nonumber\\
&&[p_{k},p_{l}]=- i\hbar\frac{\epsilon^{klm}eB_{m}}{1+e\vec{B}\cdot\vec{\Omega}}.
\end{eqnarray}
which are exact for either $B_{m}=0$ or $\Omega_{m}=0$.
These anomalous commutation relations are the main features of the action \eqref{action-1}, in contrast to the canonical commutation relations used in the Born-Oppenheimer approximation.

We summarize briefly our main assertion in this paper:
We start with a master Hamiltonian formalism and derive the action \eqref{action-1} by a second quantized path integral formulation \cite{deguchi} using the adiabatic condition $|E_{n\pm1}-E_{n}|\gg 2\pi\hbar/T$, where $T$ is the typical time scale of the slower system and $E_{n}$ is the energy level of the fast system.  In the conventional adiabatic limit, $T\rightarrow$ large with $E_{n}$ fixed, the extension of the action \eqref{action-1} to the domain of variables covering the full phase space including rapid movement is not allowed, and thus the equal-time commutation relations are not determined by the BJL prescription \footnote{ The basic idea of the BJL prescription, which is more general than the canonical quantization in the sense that it works for the system with quantum anomalies also,  is that the equal-time commutation relation of two operators $A(t)$ and $B(t)$ is determined by the analysis of the short-time limit of the time ordered product $\langle TA(t_{1})B(t_{2})\rangle$ with $t_{1}\rightarrow t_{2}$, which in turn means that the infinitely large frequencies of the Fourier transform of $\langle TA(t_{1})B(t_{2})\rangle$ determine the equal-time commutation relation.}. One may recognize the starting canonical commutation relations by going back to the starting Lagrangian. In contrast,  a singular limit of the adiabatic condition with 
$|E_{n\pm1}-E_{n}|\rightarrow \infty$ 
allows any motion of the slower variables to be regarded as adiabatic motion; we regard that this limit corresponds to the customary treatment of  \eqref{action-1}.
In this limit, however, one recognizes that one cannot quantize the action \eqref{action-1} in a canonical manner \cite{Duval}, although the action \eqref{action-1} is derived starting with a formalism defined in the conventional canonical commutation relations. It is customary to regard \eqref{action-1} as a classical action and applies a second-time quantization by applying an extended phase space formalism \cite{Faddeev-Jackiw}, for example, and one obtains the deformed Poisson brackets \eqref{Poisson bracket} or the deformed commutators \eqref{commutator-1} by a background field method \cite{fujikawa-prd2018}. The deformed phase space volume associated with the deformed Poisson brackets leads to a modified density of states $\propto 1+ e\vec{B}\cdot\vec{\Omega}$ and the Nernst effect in thermal functions \cite{Niu, Niu2}. We emphasize that \eqref{action-1} is formally invariant under both gauge transformations of the electromagnetic field and  of  Berry's phase.

In the Born-Oppenheimer approximation that is based on canonical commutation relations, one first observes that the gauge symmetry of Berry's connection is a measure of the validity of Born-Oppenheimer approximation where the total system is written as $\Psi=\sum_{l}\varphi_{l}(P)\phi_{l}(x,P)$.  The gauge symmetry of Berry's connection implies that the subsystem $\varphi_{l}(P)\phi_{l}(x,P)$, which consists of a specific energy level of the fast system $\phi_{l}(x,P)$ and the slower system $\varphi_{l}(P)$ with Berry's connection ${\cal A}^{(l)}_{k}(P)$,   is treated independently of  the total system $\Psi$. This independence of each subsystem is not compatible with the electromagnetic gauge symmetry of the vector potential $eA_{k}$, which acts on all the subsystems universally in $\Psi=\sum_{l}\varphi_{l}(P)\phi_{l}(x,P)$.  
Thus a subsystem which satisfies the gauge symmetry of Berry's connection cannot incorporate  consistently the electromagnetic potential $eA_{k}$, as we demonstrate explicitly. 
Namely, the electromagnetic gauge symmetry and  the gauge symmetry of Berry's connection are not compatible in the Born-Oppenheimer approximation which is defined by canonical commutation relations. We also discuss an algebraic manifestation of this incompatibility.
We argue that this incompatibility in the Born-Oppenheimer approximation is a restatement of the failure of the canonical quantization of the action \eqref{action-1} in the presence of the electromagnetic vector potential $eA_{k}$ and Berry's phase. This is because if the canonical quantization of the action \eqref{action-1} should be defined, it would imply that the Born-Oppenheimer approximation, which is defined by canonical commutation relations, satisfies the gauge symmetry of the electromagnetism and the gauge symmetry of Berry's connection simultaneously. We explicitly demonstrate that the Born-Oppenheimer approximation, when converted to the Lagrangian  (path integral) formalism, can incorporate the vector potential $eA_{k}(X)$ in a gauge invariant manner, but the resulting action is not amenable to the conventional canonical formulation of quantum mechanics. We conclude that the equations of motion of the anomalous Hall effect \eqref{semi-classical equation} are generally valid in the precise adiabatic limit, but the appearance of the non-canonical dynamical system and the Nernst effect is based on the deformation of the quantum principle to incorporate the two incompatible gauge symmetries.

\section{Path integral derivation }

We use the path integral formulation of Berry's phase discussed in \cite{deguchi}. We adopt the Hamiltonian of the form $H=H_{0} + H_{1}$ assuming that the slower particle is charged with $q=-e$ ($e>0$) and the fast particle is neutral for simplicity,
\begin{eqnarray}\label{starting Hamiltonian2}
H_{0}(X, P+ eA(X))&=&\frac{1}{2M}(P_{k} + eA_{k}(X))^{2} -e\phi(X),\nonumber\\
H_{1}(x, p; P+ eA(X))&=&H_{1}(x_{k}, p_{k} ; P_{k} + eA_{k}(X))
\end{eqnarray}
with the canonical quantization of fast variables
 \begin{eqnarray}\label{standard commutator-0}
 [p_{k}, x_{l}]=\frac{\hbar}{i}\delta_{kl}, \ \ [p_{k}, p_{l}]=0, \ \ [x_{k}, x_{l}]=0
 \end{eqnarray}
 and  the canonical quantization of slower variables
 \begin{eqnarray}\label{standard commutator-2}
 [P_{k}, X_{l}]=\frac{\hbar}{i}\delta_{kl}, \ \ [P_{k}, P_{l}]=0, \ \ [X_{k}, X_{l}]=0
 \end{eqnarray}
by treating $H_{0}(X, P)$ as the slower system; from now on, we use the capital characters $X_{k}$ and $P_{k}$ for slower variables without stated otherwise ( in contrast, the lower case letters are sometimes used for slower variables also to respect the past usages).  The use of covariant derivative, $P_{k} + eA_{k}(X)$,  ensures the electromagnetic gauge invariance. We discuss the case of a single fast particle and a single slow particle, for simplicity, although our use of the second quantization can cover a slightly more general case~\footnote{The general proof of the adiabatic theorem \cite{Kato} includes the case of degenerate states. We can incorporate the case with the degeneracy of $n$-states of the fast system using an internal $U(n)$ symmetry for a single particle. See also \cite{Wilczek}.}.
We start with the fundamental path integral
\begin{eqnarray}\label{starting path integral}
Z
&=&\int {\cal D}\overline{P}_{k}{\cal D}X_{k}\exp\{\frac{i}{\hbar}\int_{0}^{T} dt [(\overline{P}_{k}(t)-eA_{k}(X(t)))\dot{X}_{k}(t)
- H_{0}(X(t), \overline{P}(t))]\}\\
&\times&\int {\cal D}\psi^{\star}{\cal D}\psi\exp\{\frac{i}{\hbar}\int_{0}^{T} dt\int d^{3}x [\psi(t,\vec{x})^{\star}i\hbar \partial_{t}\psi(t,\vec{x}) - \psi(t,\vec{x})^{\star}H_{1}(\frac{\hbar}{i}\vec{\nabla}, \vec{x}; \overline{P}(t))\psi(t,\vec{x})]\} \nonumber
\end{eqnarray}
with 
\begin{eqnarray}
\overline{P}_{k}(t)=P_{k}(t) + eA_{k}(X(t))
\end{eqnarray}
where we used the invariance of the path integral measure
${\cal D}P_{k}{\cal D}X_{k}={\cal D}\overline{P}_{k}{\cal D}X_{k}$.  
We use the second quantization for the fast system, which is convenient to analyze the level crossing and Berry's phase \cite{deguchi}.
It is confirmed that the BJL prescription reproduces the canonical commutation relations  \eqref{standard commutator-2} and 
\begin{eqnarray}
[\psi(t,\vec{x}), \psi^{\dagger}(t,\vec{y})]=\delta(\vec{x}-\vec{y})
\end{eqnarray}
in the path integral formula of \eqref{starting path integral}.
We expand the field variable $\psi(t,\vec{x})$ into a complete set of orthonormal bases $\{\phi_{k}(\vec{x}; \overline{P}(t))\}$ defined by 
\begin{eqnarray}\label{instantaneous mode expansion}
&&H_{1}(\frac{\hbar}{i}\vec{\nabla}, \vec{x}; \overline{P}(t))\phi_{n}(\vec{x}; \overline{P}(t))
=E_{n}(\overline{P}(t))\phi_{n}(\vec{x}; \overline{P}(t)),\nonumber\\
&&\psi(t,\vec{x})=\sum_{n}a_{n}(t)\phi_{n}(\vec{x}; \overline{P}(t)).
\end{eqnarray}
 We then have 
\begin{eqnarray}\label{final result of second quantization}
Z
 &=&\int {\cal D}\overline{P}_{k}{\cal D}X_{k}\exp\{\frac{i}{\hbar}\int_{0}^{T} dt [(\overline{P}_{k}(t)-eA_{k}(X(t)))\dot{X}_{k}(t)
- H_{0}(X(t), \overline{P}(t))]\}\nonumber\\
&\times&\int \Pi_{n}{\cal D}a^{\star}_{n}{\cal D}a_{n}\exp\{\frac{i}{\hbar}\int_{0}^{T} dt\sum_{n}[ a_{n}^{\star}(t)i\hbar \partial_{t}a_{n}(t)- E_{n}(\overline{P}(t)) a_{n}^{\star}(t)a_{n}(t)]\}\nonumber\\
&&\hspace{2cm} \times \exp\{\frac{i}{\hbar}\int_{0}^{T} dt \sum_{n,l}\langle n,t|i\hbar\partial_{t}|l,t\rangle a_{n}^{\star}(t)a_{l}(t) \}\nonumber\\
&\simeq&\int {\cal D}\overline{P}_{k}{\cal D}X_{k}\exp\{\frac{i}{\hbar}\int_{0}^{T} dt [(\overline{P}_{k}-eA_{k}(X))\dot{X}_{k}
- H_{0}(X, \overline{P})]\}\nonumber\\
&\times&\int \Pi_{n}{\cal D}a^{\star}_{n}{\cal D}a_{n}\exp\{\frac{i}{\hbar}\int_{0}^{T} dt\sum_{n}[ a_{n}^{\star}(t)i\hbar \partial_{t}a_{n}(t)- \left(E_{n}(\overline{P}) -{\cal A}^{(n)}_{k}(\overline{P})\dot{\overline{P}^{k}}\right)a_{n}^{\star}(t)a_{n}(t)]\}\nonumber\\
\end{eqnarray} 
where we defined
\begin{eqnarray}\label{diagonal approximation}
\langle n,t|i\hbar\partial_{t}|l,t\rangle &=& \int d^{3}x\phi^{\dagger}_{n}(\vec{x}; \overline{P}(t))i\hbar\partial_{t}\phi_{l}(\vec{x}; \overline{P}(t)),\nonumber\\
{\cal A}^{(n)}_{k}(\overline{P})\dot{\overline{P}^{k}}(t) &\equiv& \langle n,t|i\hbar\partial_{t}|n,t\rangle\nonumber\\
&=& \int d^{3}x\phi^{\dagger}_{n}(\vec{x}; \overline{P}(t))i\hbar\frac{\partial}{\partial \overline{P}^{k}(t)}\phi_{n}(\vec{x}; \overline{P}(t))\dot{\overline{P}^{k}}(t).
\end{eqnarray} 
The diagonal element ${\cal A}^{(n)}_{k}(\overline{P})\dot{\overline{P}^{k}}$ is commonly called Berry's phase. The assumption of the diagonal dominance, which was used in the last step of the above path integral formula, corresponds to the {\em adiabatic approximation} and it is valid only for 
\begin{eqnarray}\label{fundamental condition}
 |E_{n\pm1}(\overline{P}) -E_{n}(\overline{P})| \gg \hbar \frac{2\pi}{T_{s}}
\end{eqnarray}
where $T_{s}$ stands for the typical time scale of the slower system; usually $T_{s}$ is estimated by the period of the  slowly varying variable $\overline{P}_{k}(t)$, $\overline{P}_{k}(0)=\overline{P}_{k}(T_{s})$. We understand that $\hbar (2\pi/T_{s})$ stands for the typical energy scale contained in $\overline{P}_{k}(t)$ \footnote{We use $T_{s}$ for the period of the slower dynamical system to distinguish it from the T-ordering operation, whenever necessary. It is convenient to choose the upper-bound of the time integral in the path integral \eqref{final result of second quantization} to agree with this period $T_{s}$.}. This condition means that the movement (or energy) of the slower system is much slower (or smaller)  than the level  spacing of the fast system.
If this adiabaticity condition is not satisfied, one needs to retain the non-diagonal transition elements $\langle n,t|i\hbar\partial_{t}|l,t\rangle$ in \eqref{final result of second quantization}, and the path integral is reduced to the starting expression \eqref{starting path integral}  after summing over $n$ and $l$.  This shows that one would recover the starting universal canonical commutation relations for the slower variables \eqref{standard commutator-2} if one evaluates the fast system exactly without using the adiabatic approximation \footnote{For the moment, we understand the condition \eqref{fundamental condition} in the manner of the standard adiabatic condition $T\rightarrow {\rm large}$ with energy eigenvalues fixed.} in the present setting. 

The slower system standing on the specific n-th level 
\begin{eqnarray}
a_{n}^{\dagger}(0)|0\rangle
\end{eqnarray}
 of the fast system, namely, the fast system being constrained to the n-th level,  is described by the path integral derived from \eqref{final result of second quantization}~\footnote{The conversion of the path integral to the time-evolution operator for the variables $\{a_{n}, a_{n}^{\dagger}\}$ is 
\begin{eqnarray}
&&\int \Pi_{n}{\cal D}a^{\star}_{n}{\cal D}a_{n}\exp\{\frac{i}{\hbar}\int_{0}^{T} dt \sum_{n} [a_{n}(t)^{\star}i\hbar \partial_{t}a_{n}(t)\nonumber\\
&&\hspace{4cm} - \left(E_{n}(\overline{P}(t)) -{\cal A}^{(n)}_{k}(\overline{P}(t))\dot{\overline{P}^{k}}(t)\right)a_{n}^{\star}(t)a_{n}(t)]\}\nonumber\\
&&\rightarrow \exp\{\sum_{n}\frac{-i}{\hbar}\int_{0}^{T}dt 
\left(E_{n}(\overline{P}(t)) -{\cal A}^{(n)}_{k}(\overline{P}(t))\dot{\overline{P}^{k}}(t)\right)a^{\dagger}_{n}(0)a_{n}(0)
\}\nonumber
\end{eqnarray}
by noting $a^{\dagger}_{n}(t)a_{n}(t)=a^{\dagger}_{n}(0)a_{n}(0)$.}
\begin{eqnarray}\label{final path integral formula}
Z_{n}
 &=&\int {\cal D}\overline{P}_{k}{\cal D}X_{k}\\
 &\times&\exp\{\frac{i}{\hbar}\int_{0}^{T} dt [(\overline{P}_{k}-eA_{k}(X))\dot{X}_{k}
- H_{0}(X, \overline{P})- (E_{n}(\overline{P}) -{\cal A}^{(n)}_{k}(\overline{P})\dot{\overline{P}_{k}})]\}\nonumber
\end{eqnarray} 
with $H_{0}(X, \overline{P})=\frac{1}{2M}\overline{P}_{k}^{2} -e\phi(X)$. This formula is valid under the crucial constraint \eqref{fundamental condition}.
The Lagrangian appearing in this path integral
\begin{eqnarray}\label{effective Lagrangian-n}
L_{n}=(\overline{P}_{k}-eA_{k}(X))\dot{X}_{k} +{\cal A}^{(n)}_{k}(\overline{P})\dot{\overline{P}_{k}}
- (\frac{1}{2M}\overline{P}_{k}^{2}+E_{n}(\overline{P})) + e\phi(X)
\end{eqnarray}
precisely  agrees with the common Lagrangian in condensed matter physics \eqref{action-1} if one identifies
\begin{eqnarray}
\overline{P}_{k}\rightarrow p_{k}, \ \ X_{k}\rightarrow x_{k},\ \ {\rm and}\ \ \epsilon_{n}(p)=\frac{1}{2M}\overline{P}_{k}^{2} + E_{n}(\overline{P}), 
 \end{eqnarray}
respectively, and if one uses the adiabatic Berry's phase ${\cal A}^{(n)}_{k}$ associated with the specific $n$-th level of the fast system.  Our path integral formulation thus naturally reproduces the common formulas \eqref{action-1} in the precise adiabatic limit but with the adiabatic Berry's phase  ${\cal A}^{(n)}_{k}$ instead of a genuine Dirac monopole. 
The action appearing in \eqref{final path integral formula} is invariant under a gauge transformation of Berry's phase 
\begin{eqnarray}\label{gauge transformation of Berry's phase}
{\cal A}^{(n)}_{k}(\overline{P}) \rightarrow {\cal A}^{(n)}_{k}(\overline{P}) + \frac{\partial}{\partial \overline{P}_{k}} \omega^{(n)}(\overline{P})
\end{eqnarray}
with different $\omega^{(n)}(\overline{P})$ for each $n$ if one chooses the periodic boundary condition $\overline{P}_{k}(0)=\overline{P}_{k}(T)$. 
The induced Berry's phase term is invariant by itself
\begin{eqnarray}\label{gauge invariance of Berry phase}
\int_{0}^{T}dt\{{\cal A}^{(n)}_{k}(\overline{P}) + \frac{\partial}{\partial \overline{P}_{k}} \omega^{(n)}(\overline{P})\}\dot{\overline{P}_{k}} &=&\int_{0}^{T}dt{\cal A}^{(n)}_{k}(\overline{P})\dot{\overline{P}_{k}} +\int_{0}^{T}dt\frac{d}{dt}\omega^{(n)}(\overline{P}(t))\nonumber\\
&=&\int_{0}^{T}dt{\cal A}^{(n)}_{k}(\overline{P})\dot{\overline{P}_{k}}. 
\end{eqnarray}
This gauge invariance is a direct consequence of the adiabatic (diagonal) approximation in \eqref{diagonal approximation}, which is based on the crucial condition \eqref{fundamental condition}. This gauge invariance is thus a measure of the validity of the adiabatic approximation.

\subsection{Equations of motion without using commutation relations}

 It is important that  the path integral \eqref{final path integral formula} with the Lagrangian \eqref{effective Lagrangian-n} is an approximation in the following two senses: Firstly, it is an approximation since all the states other than $n$ and their mixings with the state $n$ have been neglected. Secondly, to justify the above truncation, the valid energy  domain of the path integral formula (for slower variables) is limited to the very slow motions of the slower variables.  If one goes outside this energy domain, the above path integral formula with the given $L_{n}$ is not accurate. 
 
 We recall that the equations of motion are defined in the path integral without using canonical commutation relations:  We illustrate it by starting with the identity
\begin{eqnarray} \label{identity}
 &&\int {\cal D}\overline{P}_{k}{\cal D}X_{k} \exp[\frac{i}{\hbar}S(\overline{P}, X)]\nonumber\\
&=&
 \int {\cal D}(\overline{P}_{k}+\alpha_{k}){\cal D}(X_{k}+\beta_{k})  \exp[\frac{i}{\hbar}S(\overline{P}+\alpha, X+\beta)]\nonumber\\
 &=&
 \int {\cal D}\overline{P}_{k}{\cal D}X_{k} \exp[\frac{i}{\hbar}S(\overline{P}+\alpha, X+\beta)]
\end{eqnarray}
where $S=\int dt L$ and $\alpha_{k}(t)$ and $\beta_{k}(t)$ are infinitesimal arbitrary functions of $t$ but {\em independent of} $\overline{P}_{k}$ and $X_{k}$. The first equality  in \eqref{identity} is an identity since the path integral in terms of $\{\overline{P}_{k},\ X_{k}\}$ and the path integral in terms of $\{\overline{P}_{k} +\alpha_{k},\ X_{k}+\beta_{k}\}$ are the same, namely, the change of the naming of path integral variables does not change the path integral itself. The basic postulate of the path integral is that 
\begin{eqnarray}
{\cal D}(\overline{P}_{k}+\alpha_{k}){\cal D}(X_{k}+\beta_{k}) =
{\cal D}\overline{P}_{k}{\cal D}X_{k}
\end{eqnarray}
namely, no nontrivial Jacobian for this form of the change of variables, and thus we obtain the last line of \eqref{identity}. After expanding the action in the linear order of the infinitesimal $\alpha_{k}$ and $\beta_{k}$ in the last expression of \eqref{identity}, one obtains
\begin{eqnarray}\label{variational principle}
&&\langle T^{\star}\int dt^{\prime} \alpha_{k}(t^{\prime})\frac{\delta S}{\delta \overline{P}_{k}(t^{\prime})}\rangle =0, \nonumber\\
&&\langle T^{\star}\int dt^{\prime} \beta_{k}(t^{\prime})\frac{\delta S}{\delta X_{k}(t^{\prime})}\rangle=0.
\end{eqnarray}
By taking the functional derivative of these relations with respect to $\alpha_{k}(t)$ and $\beta_{k}(t)$, respectively, one obtains \eqref{semi-classical equation}
\begin{eqnarray}\label{semi-classical equation3}
&&\langle T^{\star}\{\dot{X}_{k}(t)+\Omega^{(n)}_{kl}(\overline{P})\dot{\overline{P}}_{l}(t) -\frac{\partial [\frac{1}{2M}\overline{P}_{k}^{2} + E_{n}(\overline{P})](t)}{\partial \overline{P}_{k}}\}\rangle=0, \nonumber\\
&&\langle T^{\star}\{\dot{\overline{P}}_{k}(t)+eF_{kl}(X)\dot{X}_{l}(t) - e\frac{\partial}{\partial X_{k}}\phi(X(t))\}\rangle=0,
\end{eqnarray}
with the Berry's curvature defined in the $n$-th level of the fast system
\begin{eqnarray}
\Omega^{(n)}_{kl}(\overline{P}) = \frac{\partial {\cal A}^{(n)}_{l}(\overline{P})}{\partial \overline{P}_{k}} -
\frac{\partial {\cal A}^{(n)}_{k}(\overline{P})}{\partial \overline{P}_{l}}.
\end{eqnarray}

In the present simple case, the difference between the covariant $T^{\star}$-product and the ordinary $T$-product does not appear, and thus one may replace $T^{\star}$ by $T$ in the above formula. It is important that 
the quantum equations of motion are derived without using  the commutation relations in the path integral (Lagrangian) formalism.

\subsection{Commutation relations} 

The fundamental canonical commutation relations are always valid in the precise treatment of the fast system without adiabatic approximations. This observation is natural and consistent with our basic path integral formulation \eqref{final result of second quantization}; if one sums over all the levels $n$ and $l$ of the fast system, one comes back to the starting fundamental formula \eqref{starting path integral} for which one obviously recovers the basic canonical commutation relations by the BJL procedure for the slower variables.  The basic canonical commutation relations are valid exactly in this case, which covers all the allowed energy range of the fast and slower systems. 

It is important how one incorporates the above general observation into the understanding of the path integral formula for a slower system
\begin{eqnarray}\label{final path integral formula-3}
Z_{n}
 &=&\int {\cal D}\overline{P}_{k}{\cal D}X_{k}\exp\{\frac{i}{\hbar}\int_{0}^{T} dt L_{n}\}
\end{eqnarray} 
with the Lagrangian \eqref{effective Lagrangian-n}
which  agrees with the common Lagrangian in condensed matter physics \eqref{action-1}.

When one discusses the commutation relations, one may consider two main options: \\
(i) The above effective Lagrangian is defined in the (conventional) adiabatic limit $T\rightarrow$ large and thus it is not used to quantize the slower variables by the BJL prescription, for which infinitely large frequencies are crucial.  
A natural extension of the above Lagrangian to the non-adiabatic domain, where one defines the commutation relations by the BJL procedure and thus one can estimate the plausible commutators,  is to go back to the starting well-defined path integral formula \eqref{starting path integral} which gives rise to the normal canonical commutation relations 
\begin{eqnarray}\label{standard commutator-3}
 [P_{k}, X_{l}]=\frac{\hbar}{i}\delta_{kl}, \ \ [P_{k}, P_{l}]=0, \ \ [X_{k}, X_{l}]=0
 \end{eqnarray}
with $
\overline{P}_{k}(t)=P_{k}(t) + e A_{k}(X(t))$.
This is based on the assumption that the starting Lagrangian is valid in the non-adiabatic domain as well as the adiabatic domain where one recognizes Berry's phase. This appears to be logically consistent and we call this interpretation as a standard adiabatic approximation in this article; the procedure described above shall be illustrated later in a concrete example of a Weyl fermion appearing in the two-level crossing in condensed matter physics. In this option, the path integral \eqref{final path integral formula-3} is a formal object that may be used to derive the equations of motion of the anomalous Hall effect \eqref{semi-classical equation}, which is derived without using commutation relations. The formula \eqref{semi-classical equation} is then regarded as a {\em useful but  approximate formula} valid only in the standard adiabatic condition \eqref{fundamental condition} \footnote{  In this view point, our understanding of the adiabatic  Berry's phase in the anomalous Hall effect  is analogous to that of Schwinger's anomalous magnetic moment in QED, which modifies the low-energy effective equations of motion of spin by an order $\hbar$ correction but does not modify the canonical commutation relations of the electron field.  }. In this understanding, we do not encounter the Nernst effect.\\
(ii) The second  option may be to adopt  a singular limit of the adiabatic condition \eqref{fundamental condition} defined by
\begin{eqnarray}\label{alternative adiabatic limit3}
|\epsilon_{n\pm1} - \epsilon_{n}| \rightarrow \infty 
\end{eqnarray}
with fixed $\epsilon_{n}=\frac{1}{2M}\overline{P}_{k}^{2} + E_{n}(\overline{P})$, for which any ordinary motions of 
slower variables $X_{k}(t)$ and $P_{k}(t)$ are regarded as adiabatic.
One may then 
treat the above  effective adiabatic Lagrangian \eqref{effective Lagrangian-n} to be a classical one since the commutation relations are generally modified in this singular limit.  
The Lagrangian is then extended to the non-adiabatic domain as it is and then quantized by a suitable means. Another important property is that Berry's phase $\Omega^{(n)}_{kl}(\overline{P})$  becomes generally an exact Dirac monopole in this singular limit. This is the scheme one applied to  \eqref{action-1} in the past.
One then applies a renewed (second-time) quantization by BJL or other methods to the effective Lagrangian \eqref{action-1} or equivalently to \eqref{effective Lagrangian-n} and recognizes the appearance of the non-canonical system with anomalous Poisson brackets \eqref{Poisson bracket} or anomalous commutation relations \eqref{commutator-1} in the background field method.

As for the appearance of non-canonical property (i.e., not quantized in a conventional manner)  of the Lagrangian \eqref{action-1} or \eqref{effective Lagrangian-n} for $eA_{k}\neq0$ \footnote{The canonical quantization is possible for $eA_{k}=0$, if one uses $X^{(l)}_{k}=X_{k}-{\cal A}^{(l)}_{k}$ as an auxiliary variable.}, we attribute it to the truncation of the well-defined quantum system \eqref{starting path integral} to sub-systems in which both the gauge symmetry of the electromagnetic vector potential and the newly introduced  gauge symmetry of the adiabatic Berry's phase \eqref{gauge transformation of Berry's phase} are formally preserved in the singular limit \eqref{alternative adiabatic limit3}, but they are incompatible in the Hamiltonian formalism. The algebraic compatibility shall be analyzed in subsection 3.2 later in the context of the Born-Oppenheimer approximation. 

This second option singles out the slower system constrained to the neighborhood of a specific $n$-th energy level $E_{n}$ of the fast system  by a singular limit of the adiabatic condition \eqref{alternative adiabatic limit3}. 
If one is interested in a specific $n$-th energy band and its intra-band physics such as the Nernst effect in condensed matter physics, this second option may be physically a sensible one.

\section{Born-Oppenheimer approximation and constrained dynamics}

\subsection{A manageable model}
An analysis of  a manageable  model  in  the Born-Oppenheimer approximation, which is readily compared to the scheme with Berry's phase in Section 2, shall be given in this section to illustrate that the Born-Oppenheimer approximation, which operates within the scheme of the canonical Hamiltonian formalism, does not deform the principle of quantum mechanics in the slower system, although the  auxiliary variables which satisfy the deformed commutation relations are commonly used to describe the anomalous Hall effect \cite{Blount, Sinitsyn}. 

 In the formulation of the Born-Oppenheimer approximation, one starts with the time-independent master Schr\"{o}dinger equation (using a simplified model described by $x^{k}$ and $X^{k}$, as an example),
 \begin{eqnarray}\label{BO-1}
 [H_{0}(X, P) + H_{1}(x, p; P)]\Psi(x, P)=E\Psi(x, P)
  \end{eqnarray}
 which implies that we adopt the representation of $x_{k}$ and $P_{k}$ diagonal, although the canonical formalism is basically symmetric with respect to coordinates and momenta; the possible inclusion of the vector potential $eA_{k}(X)$ and the associated complications will be discussed later. 
The canonical quantization of the fast variables is 
 \begin{eqnarray}
 [p_{k}, x_{l}]=\frac{\hbar}{i}\delta_{kl}, \ \ [p_{k}, p_{l}]=0, \ \ [x_{k}, x_{l}]=0
 \end{eqnarray}
 and the canonical quantization of the slower variables is given by
 \begin{eqnarray}\label{BO-standard commutator-2}
 [P_{k}, X_{l}]=\frac{\hbar}{i}\delta_{kl}, \ \ [P_{k}, P_{l}]=0, \ \ [X_{k}, X_{l}]=0.
 \end{eqnarray}
 The present analysis goes through for any finite $N$ number of fast coordinates $x_{k}$ without any significant modification. We however consider a single freedom for each of the fast and slower systems, for simplicity. 
  
As a first step, we  confirm that the typical quantum mechanical solutions of $H_{0}(X,P)$ generate slow motions compared to the expected motion in $H_{1}$; we thus treat $P_{k}$ as slower variables.  We then expand the total wave function
 \begin{eqnarray}\label{BO-2}
 \Psi(x, P)=\sum_{n}\varphi_{n}(P)\phi_{n}(x, P)
 \end{eqnarray}
 by solving the equation
 \begin{eqnarray}\label{BO-3}
 H_{1}(x, p; P)\phi_{n}(x, P)=E_{n}(P)\phi_{n}(x, P)
 \end{eqnarray}
with $P_{k}$ treated as background variables; the states $\{\phi_{n}(x, P) \}$ are assumed to form a complete orthonormal basis set of the fast system. By inserting \eqref{BO-2} into the equation \eqref{BO-1} and multiplying by $\phi^{\star}_{l}(x,P)$ and  integrating over $x_{k}$,  one obtains 
\begin{eqnarray}
\sum_{n}\{\int d^{3}x \phi_{l}(x, P)^{\star}H_{0}(X, P)\phi_{n}(x, P)+ E_{n}(P)\delta_{l,n}\} \varphi_{n}(P) =E\varphi_{l}(P).
\end{eqnarray}
We tentatively adopt in this subsection for simplicity
\begin{eqnarray}\label{Hamitonian of slow variables-2}
H_{0}&=&\frac{1}{2M}P_{k}^{2} +\frac{M\omega_{0}^{2}}{2}X_{k}^{2}
\end{eqnarray}
 which makes the analysis transparent without extra technical complications. We give later a non-trivial example.
Using the completeness relation $\sum_{l^{\prime}}\phi^{\star}_{l^{\prime}}(x,P)\phi_{l^{\prime}}(y,P)=\delta^{3}(x^{k}-y^{k})$, we have
\begin{eqnarray}\label{exact Born-Oppenheimer}
&&\sum_{n}\sum_{l^{\prime}}\{\frac{M\omega_{0}^{2}}{2}(\delta^{ll^{\prime}}\frac{-\hbar}{i}\nabla_{k}+{\cal A}^{ll^{\prime}}_{k}(P))(\delta^{l^{\prime}n}\frac{-\hbar}{i}\nabla_{k}+{\cal A}^{l^{\prime}n}_{k}(P)) + \frac{1}{2M}P_{k}^{2}\delta_{ln}\nonumber\\
&&\hspace{1.5cm} +E_{n}(P)\delta_{l,n}\}\varphi_{n}(P)
=E\varphi_{l}(P)
\end{eqnarray}
where $X_{k}=\frac{-\hbar}{i}\frac{\partial}{\partial P_{k}}= \frac{-\hbar}{i}\nabla_{k}$ and 
\begin{eqnarray}
{\cal A}^{ll^{\prime}}_{k}(P)=\int d^{3}x  \phi_{l}(x,P)^{\star}\frac{-\hbar}{i}\frac{\partial}{\partial P_{k}}\phi_{l^{\prime}}(x,P).
\end{eqnarray}
 If one assumes the diagonal dominance ({\em adiabatic approximation}), namely, if one assumes that the slower variables $P_{k}$ do not cause a sizable mixing of fast systems described by $\phi_{l}(x,P)$ (see \eqref{fundamental condition}), or assuming that 
 the properties of the slower system $\varphi_{l}(P)$ are well described by ignoring the effects of all the states $\phi_{l^{\prime}}(x,P)$ with $l^{\prime}\neq l$, 
 one obtains 
 \begin{eqnarray}\label{time development of slow variables}
 \{\frac{M\omega_{0}^{2}}{2}(X_{k}+{\cal A}^{(l)}_{k}(P))(X_{k}+{\cal A}^{(l)}_{k}(P)) +\frac{1}{2M}P_{k}^{2} +E_{l}(P)\}\varphi_{l}(P)
=E\varphi_{l}(P)
 \end{eqnarray}
where we defined 
 \begin{eqnarray}\label{Berry connection in BO}
 {\cal A}^{(l)}_{k}(P)\equiv {\cal A}^{ll}_{k}(P) = \int d^{3}x  \phi_{l}(x,P)^{\star}\frac{-\hbar}{i}\frac{\partial}{\partial P_{k}}\phi_{l}(x,P),
 \end{eqnarray}
  which is often called Berry's {\em connection} for the specific level $l$; ${\cal A}^{(l)}_{k}(P)$ is an order 
  $\hbar$ quantity. This is the standard formula of the Born-Oppenheimer approximation. 
 
 It is thus obvious that the slower variables $X^{k}$ are quantized in the standard manner {\em only once} as in 
\eqref{BO-standard commutator-2},
and the deformation of commutation relations  $[X^{k}(t), X^{l}(t)] \neq 0$, is not induced by Berry's connection.
The system of the $l$-th subsector \eqref{time development of slow variables} has an interesting gauge symmetry defined by the simultaneous transformations 
\begin{eqnarray}\label{state specific gauge symmetry}
&&{\cal A}^{(l)}_{k}(P)\rightarrow{\cal A}^{(l)}_{k}(P) +\hbar \partial_{k}\alpha^{(l)}(P),\nonumber\\
&&\varphi_{l}(P)\rightarrow e^{i \alpha^{(l)}(P)}\varphi_{l}(P)
\end{eqnarray}
which keeps  $\Psi_{l}(x,P)=\varphi_{l}(P)\phi_{l}(x, P)$ of each $l$-th subsector invariant for a choice of $\alpha^{(l)}(P)$ for each $l$ independently; the gauge variation of ${\cal A}^{(l)}_{k}(P)$, which is  induced by the phase change of $\phi_{l}(x,P)$, keeps $\Psi_{l}(x, P)$ invariant if 
one compensates it by the phase change  of $\varphi_{l}(P)$. This gauge symmetry is thus a measure of the independence of the specific $\Psi_{l}(x,P)$ from other sectors with $l^{\prime} \neq l$, which is a measure of the validity of the adiabatic approximation.  This gauge symmetry is manifest in the equation \eqref{time development of slow variables} if one recalls that the combination 
\begin{eqnarray}\label{covariant coordinate}
X^{(l)}_{k} = X_{k}+{\cal A}^{(l)}_{k}(P)
\end{eqnarray}
defines a {\em covariant derivative} with respect to \eqref{state specific gauge symmetry}~\cite{Blount, Nagaosa}, 
which satisfies 
\begin{eqnarray}
[X^{(l)}_{k}, X^{(l)}_{m}]=-\frac{\hbar}{i}[\partial_{k} {\cal A}^{(l)}_{m}(P) -\partial_{m} {\cal A}^{(l)}_{k}(P) ].
\end{eqnarray}
In this derivation, we used  the momentum representation where the momentum is diagonal,
$X_{k}=-\frac{\hbar}{i}\frac{\partial}{\partial P_{k}}$,
which satisfies the standard canonical commutation relations \eqref{BO-standard commutator-2}.

We now discuss the corresponding equations of motion of slower variables to compare the final formula with the one given by Berry's phase. 
The effective Hamiltonian for the slower system constructed on the $l$-th level of the fast system is given by \eqref{time development of slow variables}
\begin{eqnarray}\label{BO effective Hamiltonian}
H_{l}(P)=\frac{1}{2M}P_{k}^{2} +\frac{M\omega_{0}^{2}}{2}(X_{k}+{\cal A}^{(l)}_{k}(P))(X_{k}+{\cal A}^{(l)}_{k}(P)) + E_{l}(P)
\end{eqnarray}
with $E_{l}(P)$ arising from the $l$-th level of the fast system.
We examine the quantum mechanical equations of motion of slower variables generated by the effective Hamiltonian, which is constrained to the $l$-th level of the fast system, using the canonical commutation relations (in the Heisenberg picture)
\begin{eqnarray}\label{BO-4}
\dot{P}_{m}&=&\frac{i}{\hbar}[H_{l}, P_{m}]=-M\omega_{0}^{2}(X_{m}+{\cal A}^{(l)}_{m}(P)), \nonumber\\
\dot{X}_{m}&=&\frac{i}{\hbar}[H_{l}, X_{m}]\nonumber\\
&=& \frac{M\omega_{0}^{2}}{2}[(X_{k}+{\cal A}^{(l)}_{k}(P))\frac{\partial}{\partial P_{m}}{\cal A}^{(l)}_{k}(P) +\frac{\partial}{\partial P_{m}}{\cal A}^{(l)}_{k}(P)(X_{k}+{\cal A}^{(l)}_{k}(P))]
\nonumber\\
&&+\frac{\partial}{\partial P_{m}}[ \frac{P^{2}_{k}}{2M} + E(P)]\nonumber\\
&=&- \frac{\partial}{\partial P_{m}}{\cal A}^{(l)}_{k}(P)\dot{P}_{k}+\frac{\partial}{\partial P_{m}}[ \frac{P^{2}_{k}}{2M}+ E_{l}(P)]
\end{eqnarray}
where the last expression is valid when one ignores the possible operator ordering problem \footnote{The issue of operator ordering appears when one replaces $[(X_{k}+{\cal A}^{(l)}_{k}(P))\frac{\partial}{\partial P_{m}}{\cal A}^{(l)}_{k}(P) +\frac{\partial}{\partial P_{m}}{\cal A}^{(l)}_{k}(P)(X_{k}+{\cal A}^{(l)}_{k}(P))]$  by $2 \frac{\partial}{\partial P_{m}}{\cal A}^{(l)}_{k}(P)(X_{k}+{\cal A}^{(l)}_{k}(P))$. This kind of ordering problem is common in the operator formulation in quantum mechanics. In the present case, ${\cal A}^{(l)}_{k}(P)$ is of order $O(\hbar)$ and we are interested in the order $O(\hbar)$ corrections to the equations of motion. We can thus ignore the operator ordering issue which would give $O(\hbar^{2})$ terms.}. 
If one uses the {\em auxiliary variables} $X^{(l)}_{m}\equiv X_{m}+{\cal A}^{(l)}_{m}(P)$ in \eqref{covariant coordinate} specific to the $l$-th level,
 one  has an equivalent set of equations of motion (again by ignoring the possible operator ordering problem) \cite{Blount}
\begin{eqnarray}\label{BO-5}
\dot{P}_{m}&=&-M\omega_{0}^{2}X^{(l)}_{m}, \nonumber\\
\dot{X}^{(l)}_{m}&=& - \Omega^{(l)}_{mk}(P)\dot{P}_{k}
  +\partial_{m}[\frac{P^{2}_{k}}{2M}+ E_{l}(P)]
\end{eqnarray}
with
$\Omega^{(l)}_{mk}(P)=\frac{\partial}{\partial P_{m}}{\cal A}^{(l)}_{k}(P) - \frac{\partial}{\partial P_{k}}{\cal A}^{(l)}_{m}(P).$

If one uses the electromagnetic scalar potential $e\phi(X)$ instead of the harmonic potential, \eqref{BO-5} becomes 
\begin{eqnarray}\label{BO-6}
\dot{P}_{m}&=&e\frac{\partial}{\partial X^{(l)}_{m}}\phi(X^{(l)}), \nonumber\\
\dot{X}^{(l)}_{m}&=& - \Omega^{(l)}_{mk}(P)\dot{P}_{k}
  +\partial_{m}[\frac{P^{2}_{k}}{2M}+ E_{l}(P)]
\end{eqnarray}
The set of equations \eqref{BO-6} correspond to the common equations of the anomalous Hall effect in condensed matter physics \eqref{semi-classical equation} \cite{Blount}. 

One may notice that the auxiliary variables $X^{(l)}_{m}$ \eqref{covariant coordinate} give rise to the non-commutative geometry
\begin{eqnarray}\label{canonical commutator in l-th level}
[P_{k}, P_{l}]=0, \ \ [P_{k}, X^{(l)}_{m}]=\frac{\hbar}{i}\delta_{km}, \ \ 
[X^{(l)}_{k}, X^{(l)}_{m}]= -\frac{\hbar}{i}[\partial_{k}{\cal A}^{(l)}_{m}(P) - \partial_{m}{\cal A}^{(l)}_{k}(P)],
\end{eqnarray}
specific to the $l$-th level of the fast system, in contrast to  the starting canonical commutation relations \eqref{BO-standard commutator-2}. 
But this is {\em not} the deformation of the principle of quantum mechanics, since these commutation relations of $X^{(l)}_{m}=X_{m}+{\cal A}^{(l)}_{m}(P)$
 are dictated by the canonical commutation relations \eqref{BO-standard commutator-2} of the variables $X_{m}$ and $P_{k}$; every quantity expressed in terms of $X_{m}$ and $P_{k}$ are described by the conventional canonical commutation relations, and thus no Nernst effect in \eqref{BO-6}. 
  The appearance of noncommutative geometry in the sense of \eqref{canonical commutator in l-th level} is an artifact of the choice of the auxiliary variables $X^{(l)}_{m} $ specific to the subsystem defined by $\Psi_{l}(x,P)=\varphi_{l}(P)\phi_{l}(x, P)$.  In the present example, $X_{m} (=X^{(l)}_{m} - {\cal A}^{(l)}_{m}(P))$ stand for the proper coordinates of the total system of slower particles and exhibit no non-commutative geometry.

We next examine the electromagnetic current. The measurement of the transverse velocity as an anomalous Hall effect implies that one adopts the measured electric current 
\begin{eqnarray}\label{covariant current}
 j_{m}\equiv -e\dot{X}^{(l)}_{m} = -e\dot{X}_{m} - e\dot{{\cal A}}^{(l)}_{m} =-e\dot{X}_{m} - e\left(\frac{\partial}{\partial P_{k}}{\cal A}^{(l)}_{m}\right)\dot{P}_{k}.
 \end{eqnarray}
On the other hand, 
if one evaluates the current in the starting Hamiltonian  
 \eqref{BO-1} by gauging the vector potential one has 
\begin{eqnarray}\label{exact current}
 J_{k}&=& -\frac{\delta}{\delta A_{k}(X)}[H_{0}(X, P+eA_{k}(X)) + H_{1}(x, p; P+eA_{k}(X))]|_{A_{k}=0}\nonumber\\
&=&-e\frac{\delta}{\delta P_{k}}[H_{0}(X, P+eA_{k}(X)) + H_{1}(x, p; P+eA_{k}(X))]|_{A_{k}=0}\nonumber\\
&=&-e\frac{\delta}{\delta P_{k}}[H_{0}(X, P) + 
H_{1}(x, p; P)]\nonumber\\
&=&-e \dot{X}_{k}
\end{eqnarray}
using the equations of motion for the starting exact Hamiltonian.  Thus the exact current before one makes any approximation agrees with the time derivative of the slower coordinates. If one recalls that the total wave function is given by $\Psi(x,P)=\sum_{l}\Psi_{l}(x,P)=\sum_{l}\varphi_{l}(P)\phi_{l}(x, P)$ and one is analyzing the anomalous Hall effect in the subspace  $\Psi_{l}(x,P)=\varphi_{l}(P)\phi_{l}(x, P)$, one may expect the relation 
\begin{eqnarray}\label{fundamental}
\frac{d}{dt} \sum_{l} {\cal A}^{(l)}_{m}(P) =0
 \end{eqnarray}
 in \eqref{covariant current}.
 This fundamental property \eqref{fundamental} is ensured by \eqref{starting path integral} in the Lagrangian formalism, which states that the exact evaluation of the fast system does not generate Berry's phase, combined with the adiabatic condition \eqref{fundamental condition} which implies the diagonal dominance. In the Hamiltonian formalism one can argue the same conclusion  directly from \eqref{Berry connection in BO} after a suitable regularization \footnote{Intuitively,  if one defines the completeness relation of $\phi_{l}(x, P)$ by  $\sum_{l}\phi_{l}(x, P)\phi^{\dagger}_{l}(y, P^{\prime})=\delta(x-y)\delta_{P,P^{\prime}}$, one has $\frac{d}{dt}\sum_{l}A_{l}(P)=\frac{d}{dt}\int dx\lim_{y\rightarrow x}\delta(x-y)(1/\Delta P)[\delta_{P+\Delta P,P}-\delta_{P,P}]=0$.}.
The relation \eqref{fundamental} implies that the electric current conservation is satisfied when one considers all the bands together, although the current conservation is generally modified for each band separately in the Born-Oppenheimer approximation based on the canonical commutators, if one adopts the current \eqref{covariant current}.  This observation implies that the introduction of the electromagnetic vector potential $eA_{k}(X)$, which couples to the current,  is not straightforward in the Hamiltonian of the Born-Oppenheimer approximation.
This issue is analyzed in the next subsection.

\subsection{Gauge symmetries of electromagnetism and  Berry's phase}

The introduction of the electromagnetic vector potential $A_{k}(X)$ into the adiabatic Hamiltonian with Berry's connection does not proceed in a simple manner in the Born-Oppenheimer approximation, which is defined by canonical commutation relations. 
This complication is related to  the appearance of the non-canonical system in the common description of the anomalous Hall effect \eqref{action-1} with Berry's phase, since the Born-Oppenheimer approximation operates in a canonical Hamiltonian formalism and a natural formulation with both $A_{k}(X)$ and ${\cal A}^{(l)}_{k}(P)$ would imply a canonical system instead of the non-canonical system in \eqref{action-1}.

We comment on the origin of this complication from a point of view of algebraic consistency in the framework of the Born-Oppenheimer approximation. We start with the canonical universal coordinates of the slower system
\begin{eqnarray}\label{starting canonical algebra}
[X_{k}, X_{l}]=0, \ \ \ [X_{k}, P_{l}]=i\hbar \delta_{kl}, \ \ \
[P_{k},P_{l}]=0.
\end{eqnarray}
 We define the covariant derivative $P_{k} +eA_{k}(X)$ after gauging the electromagnetic interaction and we
define the covariant derivative $X_{k} +{\cal A}^{(n)}_{k}(P)$ after inducing Berry's connection in the Hamiltonian formalism. The gauge symmetry of Berry's connection is a measure of the validity of the adiabatic approximation by constraining the coordinates of the slower system to $\varphi_{n}(P)$. We have 
\begin{eqnarray}\label{non-local product}
&&[P_{k} +eA_{k}(X), P_{l} +eA_{l}(X)]=-i\hbar 
e(\partial_{k}A_{l}(X) -\partial_{l}A_{k}(X)), \nonumber\\
&&[X_{k} +{\cal A}^{(n)}_{k}(P), X_{l} +{\cal A}^{(n)}_{l}(P)]=i\hbar (\partial_{k}{\cal A}^{(n)}_{l}(P) -\partial_{l}{\cal A}^{(n)}_{k}(P)], \nonumber\\ 
&&[P_{k} +eA_{k}(X), X_{l} +{\cal A}^{(n)}_{l}(P) ]=-i\hbar \delta_{kl} - e[{\cal A}^{(n)}_{k}(P), A_{l}(X)] 
\end{eqnarray}
where the last relation complicates the analysis, although all the commutation relations are controlled by the original canonical commutation relations \eqref{starting canonical algebra}. 
The last relation in \eqref{non-local product} shows that   the curvature on the right-hand side is non-vanishing, which implies that the two gauge symmetries are not commuting \footnote{ Besides, it contains the term 
$[{\cal A}^{(n)}_{k}(P), eA_{l}(X)]$ which is non-local and described by a Moyal product.}.
This algebraic complication is the cause of the failure of the Born-Oppenheimer approximation, which operates in the framework of the canonical Hamiltonian formalism, in the presence of the electromagnetic potential $eA_{k}(X)$. In the Hamiltonian formalism in \eqref{BO effective Hamiltonian}, if one gauges the electromagnetic vector potential $eA_{k}(X)$ by the minimal gauge principle $P_{k}\rightarrow P_{k}+eA_{k}(X)$, one obtains 
\begin{eqnarray}
&&H_{l}
=\frac{1}{2M}(P_{k} +eA_{k}(X))^{2}\\
&&+\frac{M\omega_{0}^{2}}{2}(X_{k}+{\cal A}^{(l)}_{k}(P_{k} +eA_{k}(X)))(X_{k}+{\cal A}^{(l)}_{k}(P_{k} +eA_{k}(X))) + E_{l}(P_{k} +eA_{k}(X))\nonumber
\end{eqnarray}
but this is not consistent since $X_{k}$ in $eA_{k}(X)$ needs to be replaced by the covariant derivative $X_{k} +{\cal A}^{(l)}_{k}(P)$ in the Hamiltonian formalism, ad infinitum.

We now show that the path integral (Lagrangian formalism) is flexible and show how it realizes the electric gauge invariance by sacrificing the canonical quantization.
One may start with 
\begin{eqnarray}\label{BO Lagrangian}
 L&=& P_{k}\dot{X}_{k} - H_{l}(X, P)\nonumber\\
&=& P_{k}\dot{X}_{k} - [\frac{1}{2M}P_{k}^{2} +\frac{M\omega_{0}^{2}}{2}(X_{k}+{\cal A}^{(l)}_{k}(P))(X_{k}+{\cal A}^{(l)}_{k}(P)) + E_{l}(P)]
\end{eqnarray}
where $H_{l}(X, P)$ is the Born-Oppenheimer Hamiltonian in \eqref{BO effective Hamiltonian}. One may then define the path integral
\begin{eqnarray}\label{BO path integral1}
\int {\cal D}P{\cal D}X \exp\{\frac{i}{\hbar}\int dt[P_{k}\dot{X}_{k} - H_{l}(X, P)]\}
\end{eqnarray}
which gives rise to the canonical commutation relations if the BJL prescription is applied by assuming that $H_{l}(X, P)$ is valid for any motion of $P_{k}$ and $X_{k}$. Cf. \eqref{alternative adiabatic limit3}.
One may next rewrite this path integral as 
 \begin{eqnarray}
\int {\cal D}P{\cal D}X^{(l)} \exp\{\frac{i}{\hbar}\int dt[P_{k}\dot{X}^{(l)}_{k} + {\cal A}^{(l)}_{k}(P)\dot{P}_{k}- \left(\frac{1}{2M}P_{k}^{2} +\frac{M\omega_{0}^{2}}{2}(X^{(l)}_{k})^{2} + E_{l}(P)\right)]\}
\end{eqnarray}
where $X^{(l)}_{k}=X_{k}+{\cal A}^{(l)}_{k}(P)$ and we used ${\cal D}P{\cal D}X={\cal D}P{\cal D}X^{(l)}$. The BJL prescription when applied to this path integral gives rise to canonical $[X_{k}, X_{l}]=0$, $[P_{k}, P_{l}]=0$ and $[X_{k}, P_{l}]=i\hbar\delta_{kl}$, although with non-commutative geometry $[X^{(l)}_{k}, X^{(l)}_{n}]=-\frac{\hbar}{i}\Omega^{(l)}_{kn}$ for the auxiliary variable $X^{(l)}_{k}$. One may add an electromagnetic gauge invariant term $-\int dt eA_{k}(X^{(l)})\dot{X}^{(l)}_{k}$ to the action by hand, then one obtains 
 \begin{eqnarray}\label{gauge invariant BO}
&&\int {\cal D}P{\cal D}X^{(l)} \exp\{\frac{i}{\hbar}\int dt[P_{k}\dot{X}^{(l)}_{k} - eA_{k}(X^{(l)})\dot{X}^{(l)}_{k} + {\cal A}^{(l)}_{k}(P)\dot{P}_{k} \nonumber\\
&& \hspace{3.5 cm} - \left(\frac{1}{2M}P_{k}^{2} +\frac{M\omega_{0}^{2}}{2}(X^{(l)}_{k})^{2} + E_{l}(P)\right)]\}.
\end{eqnarray}
The above extra electromagnetic coupling would be  induced from the minimally coupled Hamiltonian $$\left(\frac{1}{2M}(P_{k}+eA_{k}(X^{(l)}))^{2} +\frac{M\omega_{0}^{2}}{2}(X^{(l)}_{k})^{2} + E_{l}(P_{k}+ eA_{k}(X^{(l)}))\right)$$ by a change of the variable $P_{k}+eA_{k}(X^{(l)})\rightarrow P_{k}$ if the term ${\cal A}^{(l)}_{k}(P)\dot{P}_{k}$ should be absent, but in reality because of ${\cal A}^{(l)}_{k}(P)\dot{P}_{k}$ the manipulation in Hamiltonian formalism does not work. Note that the current to which the vector potential $eA_{k}(X^{(l)})$ at the point $X^{(l)}$ couples is chosen as the current $\dot{X}^{(l)}_{k}$. Cf. \eqref{covariant current}. 
The term $-\int dt eA_{k}(X^{(l)})\dot{X}^{(l)}_{k}$ is analogous to the Wess-Zumino term by satisfying the gauge invariance by itself in the Lagrangian formalism.  See also \eqref{gauge invariance of Berry phase}.
The Lagrangian in \eqref{gauge invariant BO}, which is confirmed not to be quantized in a canonical manner, agrees with \eqref{action-1} or \eqref{effective Lagrangian-n} if one changes the dummy path integral variable $X^{(l)}_{k}\rightarrow X_{k}$ and replaces the harmonic potential by the scalar potential $\frac{M\omega_{0}^{2}}{2}(X_{k})^{2}\rightarrow -e\phi(X_{k})$. 
This analysis shows how the Lagrangian formalism maintains the electromagnetic gauge invariance in the presence of $eA_{k}$ by sacrificing the canonical formalism of quantum mechanics.

The Nernst effect is attributed to the non-canonical behavior and a modification of the density of states $D=(1+e\vec{B}\cdot\vec{\Omega})$
\cite{Niu}, which is derived from the deformed commutators \eqref{Poisson bracket} \cite{Niu2, Sinitsyn}; the same result is obtained by a more general treatment of the action \eqref{action-1} \cite{Duval} in a singular adiabatic limit \eqref{alternative adiabatic limit3}, namely, for any ordinary movements of slower variables in \eqref{action-1}. From a point of view of gauge symmetries, the Nernst effect is regarded as a manifestation of the stricture of the electromagnetic gauge symmetry when the slower freedom is constrained to a specific $n$-th state $\varphi_{n}(P)$ in \eqref{BO-2} by imposing the gauge invariance of Berry's connection, as we have seen in the framework of the Born-Oppenheimer approximation defined by canonical commutators\footnote{In the formulation in Section 2, the singular limit of the adiabatic condition \eqref{alternative adiabatic limit3} effectively constrains the system to a specific state with $E_{n}$.}. Note that the breaking of the gauge symmetry of Berry's connection would imply that the adiabatic approximation is not accurate, while the breaking of the electromagnetic gauge symmetry would imply the inconsistency of the theory.

\section{Discussions and conclusion}

We attempted to understand why Berry's phase defined in the adiabatic limit leads to a non-canonical dynamical system with deformed commutators that have been opened up by the applications of Berry's phase to the anomalous Hall effect. We start with the derivation of  the action \eqref{action-1} or \eqref{effective Lagrangian-n} in the path integral (Lagrangian) formalism  based on the generic adiabatic condition
\begin{eqnarray}\label{adiabatic conditionS6}
|E_{n\pm 1}-E_{n}|\gg \hbar \frac{2\pi}{T}
\end{eqnarray}
where $T$ stands for the typical time scale of the slower system $X_{k}$, and $E_{n}$ is the $n$-th energy level of the fast system. Regardless of the specific interpretations of the condition \eqref{adiabatic conditionS6}, one can derive the equations of motion of the anomalous Hall effect \eqref{semi-classical equation}.  In the conventional understanding of the adiabatic condition by letting  $T\rightarrow$ large with fixed energy eigenvalues $E_{n}$ of the fast system, the BJL prescription is not applicable. But the conventional canonical commutation relations of slower variables are reproduced by applying the BJL prescription to the original exact Lagrangian, which works for the adiabatic as well as  non-adiabatic motions. The same conclusion is attained 
using an exactly solvable model of Berry's phase \cite{DF-Ann-Phys-2020} which covers from the adiabatic domain to the non-adiabatic domain. 
In this understanding of Berry's phase, no Nernst effect arises. 

On the other hand, if one considers a singular limit of the adiabatic condition 
\begin{eqnarray}\label{singular limitS6}
|E_{n\pm 1}-E_{n}|\rightarrow \infty
\end{eqnarray}
with $E_{n}$ fixed, the effective adiabatic Lagrangian is generally converted to be a classical one (since the commutation relations are generally modified in this limit ) and any ordinary motion of the slower variables satisfies the adiabatic condition \eqref{adiabatic conditionS6}. This singular limit changes the adiabatic Lagrangian, such as  \eqref{action-1} and \eqref{effective Lagrangian-n}, to a dynamical system which is not quantized in a canonical manner in the presence of  the vector potential $eA_{k}(X)$. One thus arrives at the deformed commutation relations such as 
\eqref{Poisson bracket} and \eqref{commutator-1}, which in turn leads to the Nernst effect. In the absence of the vector potential $eA_{k}(X)=0$, one still encounters the non-commutative geometry but dynamically one has a canonical system and thus no Nernst effect.

 In the Born-Oppenheimer approximation, which is defined in a Hamiltonian formalism with canonical commutation relations, one can derive the equations of motion of the anomalous Hall effect \eqref{semi-classical equation} in the absence of the vector potential $eA_{k}(X)=0$ if  one introduces the auxiliary variables 
\begin{eqnarray}\label{local coordinateS6_2}
X^{(n)}_{k}=X_{k} + {\cal A}^{(n)}_{k}(P),
\end{eqnarray}
but no Nernst effect. In the Hamiltonian formulation of the Born-Oppenheimer approximation, one cannot incorporate the vector potential $eA_{k}(X)$ in a consistent manner. We attribute this failure of the Born-Oppenheimer approximation in the presence of $eA_{k}(X)$ to the incompatibility of the gauge symmetry of Berry's connection, which projects the total system of $ \Psi(x, P)=\sum_{n}\varphi_{n}(P)\phi_{n}(x, P)$ to a specific subsystem $\varphi_{n}(P)\phi_{n}(x, P)$, and the electromagnetic gauge symmetry which acts universally over all the subsystems of $\Psi(x, P)$.  Note that the variable \eqref{local coordinateS6_2} is a covariant derivative for the subsystem $\varphi_{n}(P)\phi_{n}(x, P)$ of this gauge symmetry of Berry's connection.
We explicitly demonstrated that the Born-Oppenheimer approximation, when converted to the Lagrangian  (path integral) formalism, can incorporate the vector potential $eA_{k}(X)$ in a gauge invariant manner, but the resulting action is not amenable to the conventional canonical formulation of quantum mechanics.
 
 We thus conclude that the equations of motion of the anomalous Hall effect \eqref{semi-classical equation} are generally valid in the adiabatic limit, but the interpretation of the dynamical system represented by the action \eqref{action-1} is more involved. The appearance of the non-canonical dynamical system and the Nernst effect is based on the deformation of the quantum principle to incorporate the two incompatible gauge symmetries of  electromagnetic vector potential and Berry's connection.

From a point of view of practical applications,  
if one is interested in a specific energy band and its intra-band physics in condensed matter physics, the choice of the singular limit of the adiabatic condition \eqref{singular limitS6} may be a sensible one. The constrained dynamics of the slower system to the specific $n$-th state imposed by a singular limit of the adiabatic condition combined with the vector potential $eA_{k}(X)$ is then responsible for the deformation of the canonical formalism which leads to  the deformed commutators and the Nernst effect in the adiabatic action of the anomalous Hall effect. 

\section*{Acknowledgements}
We thank Naoto Nagaosa,  Shinichi Deguchi and Kenji Fukushima for helpful discussions and suggestions.
The present work is supported in part by JSPS KAKENHI (Grant No.18K03633).

\appendix

\section{Two-band model and commutation relations}

We illustrate that the general analysis in Section 2 works in the simple Weyl model with an explicit form of Berry phase. The model is defined by 
\begin{eqnarray}\label{level crossing Hamiltonian}
H=-\mu \vec{p}(t)\cdot \vec{\sigma}
\end{eqnarray}
where $\vec{\sigma}$ stands for the pseudo spin that describes the upper and lower crossing bands.

We thus start with  the Schr\"{o}dinger equation 
$i\hbar \partial_{t}\psi(t)=H\psi(t)$
 or the Lagrangian (in the spirit of the second quantization) given by ~\cite{Stone}
\begin{eqnarray}\label{Lagrangian}
L=\psi^{\dagger}(t)[i\hbar \partial_{t}-H]\psi(t)
\end{eqnarray}
where the two-component spinor $\psi(t)$ specifies the movement of upper and lower levels which appear in the band-crossing problem. In the present context, the {\em fast variables} are given by the spin freedom $\psi(t)$ characterized by the energy scale $\mu|\vec{p}(t)|$, and the slower variables are given by the angular freedom of $\vec{p}(t)$. For simplicity, we assume the magnitude $|\vec{p}(t)|$ to be time independent.
We then perform
a time-dependent unitary transformation $
\psi(t)= U(\vec{p}(t))\psi^{\prime}(t),\ \ 
\psi^{\dagger}(t)={\psi^{\prime}}^{\dagger}(t) 
U^{\dagger}(\vec{p}(t))$
with
$
U(\vec{p}(t))^{\dagger}\mu\vec{p}(t)\cdot\vec{\sigma}
U(\vec{p}(t))
=\mu|\vec{p}|\sigma_{3}$.
This unitary transformation is explicitly given by a $2\times2$ matrix
$U(\vec{p}(t))=\left(
             v_{+}(\vec{p})\  v_{-}(\vec{p}) \right)$,
where
\begin{eqnarray}
v_{+}(\vec{p})=\left(\begin{array}{c}
            \cos\frac{\theta}{2}e^{-i\varphi}\\
            \sin\frac{\theta}{2}
            \end{array}\right), \ \ \ 
v_{-}(\vec{p})=\left(\begin{array}{c}
            \sin\frac{\theta}{2}e^{-i\varphi}\\
            -\cos\frac{\theta}{2}
            \end{array}\right)
\end{eqnarray}
which correspond to the  use of {\em instantaneous eigenfunctions} of the operator $\mu\vec {p}(t)\cdot\vec{\sigma}$, namely, $\mu\vec {p}(t)\cdot\vec{\sigma}v_{\pm}(\vec{p})=\pm \mu|\vec{p}|v_{\pm}(\vec{p})$, where
$\vec {p}(t)=|\vec{p}|(\sin\theta\cos\varphi, \sin\theta\sin\varphi, \cos\theta)$
 with time dependent $\theta(t)$ and $\varphi(t)$.

Based on this transformation, 
 the equivalence of two 
Lagrangians is derived, namely, $L$ in \eqref{Lagrangian}
and 
\begin{eqnarray}\label{Lagrangian2}
L^{\prime}=
{\psi^{\prime}}^{\dagger}[i\hbar\partial_{t}
+\mu|\vec{p}|\sigma_{3}+U(\vec{p}(t))^{\dagger}
i\hbar\partial_{t}U(\vec{p}(t))]\psi^{\prime}.
\end{eqnarray}
 The starting Hamiltonian \eqref{level crossing Hamiltonian} is thus replaced by  \cite{Stone}
\begin{eqnarray}\label{geometric-phase}
H^{\prime}(t)&=&
-\mu|\vec{p}|\sigma_{3}- U(\vec{p}(t))^{\dagger}
i\hbar\partial_{t}U(\vec{p}(t))\nonumber\\
&=& -\mu|\vec{p}|\sigma_{3} -\hbar\left(\begin{array}{cc}
\frac{(1+\cos\theta)\dot{\varphi}}{2}&\frac{\dot{\varphi}\sin\theta+i\dot{\theta}}{2}\\
            \frac{\dot{\varphi}\sin\theta-i\dot{\theta}}{2}&
\frac{(1-\cos\theta)\dot{\varphi}}{2}
            \end{array}\right).
\end{eqnarray}
We note that  $2\mu|\vec{p}|=|\epsilon_{1}-\epsilon_{2}|$ in the notation of \eqref{fundamental condition} in Section 2.
The adiabatic condition \eqref{fundamental condition} is thus written
\begin{eqnarray}\label{adiabatic condition in band-crossing}
2\mu|\vec{p}| \gg 2\pi\hbar/T,
\end{eqnarray}
where $T$ is the typical time scale of the slower dynamical variable and customarily taken as the period of the slower dynamical
variable (i.e., angular freedom) of $\vec{p}(t)$, and we estimate $\dot{\varphi} \sim 2\pi/T$.
 We thus have from \eqref{geometric-phase}
\begin{eqnarray}\label{adiabatic Stone phase}
H^{\prime}_{ad} \simeq  -\mu|\vec{p}|\sigma_{3} -\hbar\left(\begin{array}{cc}
\frac{(1+\cos\theta)\dot{\varphi}}{2}&0\\
            0&
\frac{(1-\cos\theta)\dot{\varphi}}{2}
            \end{array}\right).
\end{eqnarray}
Berry's phase is given by the second term and one recognizes the familiar monopole-like expression in the diagonal \footnote{Berry's phase in a realistic two-band model in condensed matter physics in the adiabatic limit has been analyzed in detail by Nagaosa \cite{Nagaosa}.}.

The BJL prescription, whose essence is illustrated in Appendix B, requires the condition 
\begin{eqnarray}\label{non-adiabatic condition}
\mu|\vec{p}| \ll 2\pi\hbar/T = \hbar\omega
\end{eqnarray}
for the fixed $\mu|\vec{p}|$ to analyze the equal-time commutation relations of {\em slower variables}. 
To see the implications of the non-adiabatic condition \eqref{non-adiabatic condition} explicitly, it is convenient to perform a further unitary transformation of the 
fermionic  variable
$
\psi^{\prime}(t)= U(\theta(t))\psi^{\prime\prime}(t), \ \ \  
{\psi^{\prime}(t)}^{\dagger}=
{\psi^{\prime\prime}}^{\dagger}(t) 
U^{\dagger}(\theta(t))
$
with~\cite{deguchi}
\begin{eqnarray}\label{unitary2}
U(\theta(t))=\left(\begin{array}{cc}
            \cos\frac{\theta}{2}&-\sin\frac{\theta}{2}\\
            \sin\frac{\theta}{2}&\cos\frac{\theta}{2}
            \end{array}\right)
\end{eqnarray}
in addition to \eqref{Lagrangian2}, which diagonalizes the dominant Berry's phase term.
The Hamiltonian \eqref{geometric-phase} then becomes
\begin{eqnarray}\label{exact non-adiabatic}
H^{\prime\prime}(t)
&=&
-\mu|\vec{p}| U(\theta(t))^{\dagger}
\sigma_{3}U(\theta(t))
+ (U(\vec{p}(t))U(\theta(t)))^{\dagger}
\frac{\hbar}{i}\partial_{t}(U(\vec{p}(t))U(\theta(t)))
\nonumber\\
&=&
-\mu|\vec{p}| \left(\begin{array}{cc}
            \cos\theta&-\sin\theta\\
            -\sin\theta&-\cos\theta
            \end{array}\right)
-\hbar\left(\begin{array}{cc}
            \dot{\varphi}&0\\
            0&0
            \end{array}\right).
\end{eqnarray}
Note that the first term is bounded by $\mu|\vec{p}|$ and the second term is dominant for the non-adiabatic case $\mu|\vec{p}| \ll 2\pi\hbar/T$. We emphasize that both \eqref{geometric-phase} and \eqref{exact non-adiabatic} are exact expressions.  

The relation \eqref{exact non-adiabatic} shows that Berry's phase becomes trivial for the non-adiabatic limit \eqref{non-adiabatic condition} with fixed $\mu|\vec{p}|$ and thus no modification of commutation relations by Berry's phase  \footnote{This property has been argued to be the case  in \cite{DF-Ann-Phys-2020} using an exactly solvable model of Berry's model \cite{Fujikawa-exact}.}. 

On the other hand, if one considers a singular limit 
$
2\mu|\vec{p}|=|\epsilon_{2} -\epsilon_{1}|\rightarrow \infty
$
in the adiabatic criterion 
$2\mu|\vec{p}| \gg 2\pi\hbar/T$ in \eqref{adiabatic condition in band-crossing},
the adiabatic formula \eqref{adiabatic Stone phase} becomes exact together with an exact Dirac monopole for any movement of slower angular variables. This condition with the present two-band model leads to the action \eqref{action-1} with $A_{k} =0$ and $\phi=0$. One would thus obtain the non-commutative geometry \eqref{commutator-1} 
\begin{eqnarray}
[X_{k}, X_{l}] =i\hbar \Omega_{kl}.
\end{eqnarray}

We next show the absence of anomalous commutation relations induced by Berry's phase in the present two-level crossing model if one analyzes the exact effective Hamiltonian without recourse to adiabatic approximations. We first rewrite the exact formula \eqref{geometric-phase} as 
\begin{eqnarray}\label{geometric-phase2}
H^{\prime}(t)
= -\mu|\vec{p}|\sigma_{3} -\hbar\left(\begin{array}{cc}
\frac{(1+\cos\theta)}{2}&\frac{\sin\theta}{2}\\
            \frac{\sin\theta}{2}&
\frac{(1-\cos\theta)}{2}
            \end{array}\right)\dot{\varphi}
            -\hbar\left(\begin{array}{cc}
0&\frac{i}{2}\\
\frac{-i}{2}&0
            \end{array}\right)\dot{\theta},
\end{eqnarray}
which shows that $\dot{\theta}$ has no non-trivial conjugate variable in the phase space. It vanishes when one integrates $\int dt H^{\prime}(t)$ to define the action. The variable $\dot{\varphi}$ appears to have  non-trivial conjugate variables, which are Berry's phases and their off-diagonal partners, and thus could contribute to the modification of commutation relations.  But one can confirm that the matrix multiplying $\dot{\varphi}$ has a vanishing determinant and a unit trace, 
\begin{eqnarray}
{\rm det}\left(\begin{array}{cc}
\frac{(1+\cos\theta)}{2}&\frac{\sin\theta}{2}\\
            \frac{\sin\theta}{2}&
\frac{(1-\cos\theta)}{2}
            \end{array}\right)=0, \ \ \ 
            {\rm Tr}\left(\begin{array}{cc}
\frac{(1+\cos\theta)}{2}&\frac{\sin\theta}{2}\\
            \frac{\sin\theta}{2}&
\frac{(1-\cos\theta)}{2}
            \end{array}\right)=1
\end{eqnarray}
independently of $\theta$, which imply the eigenvalues $1$ and $0$ and thus dynamically trivial. 
To see this fact explicitly,  we refer to the unitary equivalent \eqref{exact non-adiabatic}.
 We emphasize that both \eqref{geometric-phase2} and \eqref{exact non-adiabatic} are exact expressions in the scheme of \cite{Stone}; \eqref{exact non-adiabatic} is accurate in the non-adiabatic limit with the time derivative term being diagonal.

The unitary equivalent exact \eqref{exact non-adiabatic}
shows that $\dot{\varphi}$ which multiplies would-be Berry's phases and their off-diagonal partners,  does not have any non-trivial conjugate variable, and thus does not contribute to the modification of commutation relations. The first term in \eqref{exact non-adiabatic} is essentially the same as the starting expression \eqref{level crossing Hamiltonian} if one considers the azimuthally symmetric configurations. 
Thus would-be Berry's phases and their off-diagonal partners in a precise treatment  generate no extra non-trivial  time-derivative terms in the action, and thus gives no anomalous commutation relations.

This fact agrees with the observation made in connection with \eqref{final result of second quantization} in Section 2.
The second quantized part in \eqref{final result of second quantization} except for the time derivative terms $a_{n}^{\star}(t)i\hbar \partial_{t}a_{n}(t)$ corresponds to the present 
 \eqref{exact non-adiabatic}. If one sums over $n$ and $l$ in \eqref{final result of second quantization},
the path integral is reduced to the original path integral \eqref{starting path integral} without any extra time derivative terms added, for which the canonical commutation relations of slower variables 
are recovered by the BJL prescription.
This analysis shows that the anomalous commutation relations of slower variables do not appear if no  adiabatic approximations are made in the sector of fast variables \footnote{ To be precise, there is a difference between \eqref{geometric-phase2} and \eqref{final result of second quantization} in Section 2. In \eqref{final result of second quantization}, the fast system and slower system are clearly separated, while in the Weyl model \eqref{geometric-phase2} this separation is not perfect as is seen in \eqref{exact non-adiabatic}. Thus we have no Berry's phase for the exact evaluation of the fast system in \eqref{final result of second quantization}, but we have a non-trivial Berry's phase for an exactly solvable model \cite{Fujikawa-exact, DF1} which is close to the Weyl model. The behavior of Berry's phase at the non-adiabatic limit is, however, similar in either case and becomes trivial as in \eqref{exact non-adiabatic} which is relevant when one analyzes the possible deformation of commutation relations by the BJL prescription.}.

\section{BJL analysis of commutation relations}
The basic idea of the BJL prescription \cite{BJL} is that the equal-time commutator of operators $A(t)$ and $B(t)$ is determined by the short-time limit of the correlation function $\langle T A(t_{1})B(t_{2})\rangle$ with $t_{1}\rightarrow t_{2}$, which implies the large frequency limit of the Fourier transform of $\langle T A(t_{1})B(t_{2})\rangle$.
It is illustrated how the BJL  prescription works in the present context for the simple case with  $eA_{k}(X(t))=0$.
We start with the path integral \eqref{final path integral formula} but with $eA_{k}(X(t))=0$, for which the path integral is well-defined quantum mechanically. We assume that the adiabatic Lagrangian is extended to the non-adiabatic domain of slower variables (by considering a singular limit of the adiabatic condition \eqref{singular limitS6}) 
\begin{eqnarray}\label{final path integral formula-6}
Z_{n}
 &=&\int {\cal D}\overline{P}_{k}{\cal D}X_{k}\\
 &\times&\exp\{\frac{i}{\hbar}\int_{0}^{T} dt [\overline{P}_{k}(t)\dot{X}_{k}(t) + {\cal A}_{k}(\overline{P}(t))\dot{\overline{P}_{k}}(t)
- (\frac{1}{2M}\overline{P}_{k}^{2}+E_{n}(\overline{P})) + e\phi(X)]\}.\nonumber
\end{eqnarray} 
To sketch a BJL prescription,
we analyze a simplified system instead of \eqref{final path integral formula-6},
\begin{eqnarray}
L=\overline{P}_{k}\dot{X}_{k} + {\cal A}_{k}(\overline{P})\dot{\overline{P}_{k}}
\end{eqnarray}
by ignoring the terms which do not modify the commutation relations. 
We start with the free part $L_{0}= \overline{P}_{k}\dot{X}_{k}$ which gives \eqref{standard commutator-2}; the starting correlation functions  are defined by \footnote{Feynman's $i\epsilon$ prescription is generalized by $\epsilon=\epsilon_{1} + i\mu^{2}$ using two small positive parameters $\epsilon_{1}$ and $\mu^{2}$ to define the correlation function precisely.}
\begin{eqnarray}\label{correlation functions-4}
&&\langle T X^{k}(t)\overline{P}^{l}(t^{\prime})\rangle = \int\frac{d\omega}{2\pi} e^{i\omega(t-t^{\prime})}\frac{\hbar\omega}{\omega^{2} +i\epsilon}\delta_{kl}, \nonumber\\
&&\langle TX^{k}(t)X^{l}(t^{\prime})\rangle=0, \ \  \langle T\overline{P}^{k}(t)\overline{P}^{l}(t^{\prime})\rangle=0.
\end{eqnarray}
We use the conventional $T$ product instead of the covariant $T^{\star}$ product since they agree in the present simple cases.
By multiplying the Fourier transform of the first relation $\int dt e^{-i\omega(t-t^{\prime})}\langle T X^{k}(t)\overline{P}^{l}(t^{\prime})\rangle =\frac{\hbar\omega}{\omega^{2} +i\epsilon}\delta_{kl}$ by $\omega$ and taking the limit $\omega\rightarrow \infty$, one obtains  the standard commutation relations $[X^{k}(t),\overline{P}^{l}(t)]=i\hbar \delta_{kl}$. Similarly,  $[X_{k}, X_{l}]=[\overline{P}^{k}, \overline{P}^{l}]=0$ by the BJL prescription.

We next use the term containing Berry's phase $L^{\prime}={\cal A}_{k}(\overline{P})\dot{\overline{P}_{k}}$
as a perturbation.  One then obtains the correction term by the perturbation
\begin{eqnarray}\label{perturbation by Berry's phase2}
&&\langle TX^{k}(t)X^{l}(t^{\prime})\rangle_{(1)} = \int {\cal D}\overline{P}_{k}{\cal D}X_{k}\{ X^{k}(t)X^{l}(t^{\prime})\}\nonumber\\
 &&\hspace{3cm}\times\exp\{\frac{i}{\hbar}\int_{0}^{T} dt [\overline{P}_{k}(t)\dot{X}_{k}(t)
+{\cal A}_{k}(\overline{P})\dot{\overline{P}_{k}}]\}\nonumber\\ 
&=&\int dt_{1}dt_{2}\langle TX^{k}(t)\overline{P}^{k^{\prime}}(t_{1})\rangle \{\frac{\delta}{\delta  \overline{P}^{k^{\prime}}(t_{1})}\frac{\delta}{\delta  \overline{P}^{l^{\prime}}(t_{2})}(\frac{i}{\hbar})
\int ds L^{\prime}(s)\} \langle T\overline{P}^{l^{\prime}}(t_{2})X^{l}(t^{\prime})\rangle\nonumber\\
&=&\int ds\langle TX^{k}(t)\overline{P}^{k^{\prime}}(s)\rangle(\frac{i}{\hbar})
\Omega_{k^{\prime}l^{\prime}}(s) \langle T \dot{\overline{P}}^{l^{\prime}}(s)X^{l}(t^{\prime})\rangle
\end{eqnarray}
where the curvature  induced by Berry's phase is written in the form $\Omega_{k^{\prime}l^{\prime}}(s)$.
 Using \eqref{correlation functions-4}, we thus obtain
\begin{eqnarray}\label{evaluation of commutator1}
\int dt e^{-i\omega (t-t^{\prime})}\langle TX^{k}(t)X^{l}(t^{\prime})\rangle_{(1)}
= \frac{\hbar\omega}{\omega^{2} +i\epsilon}(\frac{i}{\hbar}) \Omega_{kl}(\overline{P}(t^{\prime}))(-i\hbar).
\end{eqnarray}
Taking the limit $\omega\rightarrow \infty$ after multiplying the both hand sides by $\omega$,  one obtains the non-commutative geometry \cite{Niu, Duval}
\begin{eqnarray}\label{non-commutative geometryS3}
[X^{k}(t), X^{l}(t)] =-\frac{\hbar}{i}\Omega_{kl}(\overline{P}(t)) .
\end{eqnarray}

It has been argued in \cite{DF-Ann-Phys-2020} that \eqref{non-commutative geometryS3} is reduced to an ordinary canonical commutation relation, 
$[X^{k}(t), X^{l}(t)] =0$, if one uses 
 an exactly solvable model of Berry's phase \cite{Fujikawa-exact, DF1} in the evaluation \eqref{evaluation of commutator1}.

\end{document}